\documentclass[reqno,a4paper,final,12pt]{amsart}
\usepackage[utf8x]{inputenc}
\usepackage{textcomp}
\usepackage[a4paper,hmargin=2cm,tmargin=3cm,bmargin=3cm]{geometry}
\usepackage{amssymb,amstext,amscd,amsthm,eucal}
\usepackage{color}
\usepackage{epic}
\usepackage[pdftex]{graphicx}
\usepackage{array}
\usepackage[all]{xy}
\usepackage{hyperref}
\hypersetup{%
  pdftitle   = {The homogeneity conjecture for supergravity backgrounds},
  pdfkeywords = {supergravity, supersymmetry, superalgebra, Killing
    spinors, Lie algebras, E8, F4, homogeneity, Killing superalgebra,
    cones, holonomy},
  pdfauthor  = {José Miguel Figueroa-O'Farrill},
  pdfcreator = {\LaTeX\ with package \flqq hyperref\frqq}
}
\PrerenderUnicode{éÉ}
\newcommand{\half}{\tfrac12}

\newcommand{\fg}{\mathfrak{g}}
\newcommand{\fk}{\mathfrak{k}}
\newcommand{\fm}{\mathfrak{m}}
\newcommand{\fso}{\mathfrak{so}}

\newcommand{\Cl}{\mathrm{C}\ell}

\newcommand{\fgl}{\mathfrak{gl}}
\newcommand{\fsl}{\mathfrak{sl}}
\newcommand{\fosp}{\mathfrak{osp}}
\newcommand{\Spin}{\mathrm{Spin}}
\newcommand{\SO}{\mathrm{SO}}
\newcommand{\GL}{\mathrm{GL}}
\renewcommand{\O}{\mathrm{O}}
\ifx\Sp\UnDeFiNeD
\newcommand{\Sp}{\mathrm{Sp}}
\else
\renewcommand{\Sp}{\mathrm{Sp}}
\fi
\newcommand{\SU}{\mathrm{SU}}
\newcommand{\U}{\mathrm{U}}
\newcommand{\RR}{\mathbb{R}}
\newcommand{\CC}{\mathbb{C}}
\newcommand{\KK}{\mathbb{K}}
\newcommand{\HH}{\mathbb{H}}
\newcommand{\OO}{\mathbb{O}}
\newcommand{\PP}{\mathbb{P}}
\newcommand{\ZZ}{\mathbb{Z}}
\newcommand{\eL}{\mathcal{L}}
\newcommand{\eX}{\mathcal{X}}

\DeclareMathOperator{\Hom}{Hom}
\DeclareMathOperator{\Ric}{Ric}

\DeclareMathOperator{\dvol}{dvol}

\theoremstyle{plain}
\newtheorem{lemma}{Lemma}
\newtheorem{proposition}[lemma]{Proposition}

\newtheorem*{HomConj}{Homogeneity Conjecture}
\theoremstyle{definition}

\newcommand{\MUNCH}[1]{\relax}

\allowdisplaybreaks[1]
\begin{document}
\title{The homogeneity conjecture for supergravity backgrounds}
\author{José Miguel Figueroa-O'Farrill}
\address{Maxwell Institute and School of Mathematics, University of Edinburgh}
\address{Departament de Física Teòrica, Universitat de València}
\email{j.m.figueroa@ed.ac.uk}
\date{\today}
\begin{abstract}
  These notes record three lectures given at the workshop ``Higher
  symmetries in Physics'', held at the Universidad Complutense de
  Madrid in November 2008.  In them we explain how to construct a Lie
  (super)algebra associated to a spin manifold, perhaps with extra
  geometric data, and a notion of privileged spinors.  The typical
  examples are supersymmetric supergravity backgrounds; although there
  are more classical instances of this construction.  We focus on two
  results: the geometric constructions of compact real forms of the
  simple Lie algebras of type $B_4$, $F_4$ and $E_8$ from $S^7$, $S^8$
  and $S^{15}$, respectively; and the construction of the Killing
  superalgebra of eleven-dimensional supergravity backgrounds.  As an
  application of this latter construction we show that supersymmetric
  supergravity backgrounds with enough supersymmetry are necessarily
  locally homogeneous.
\end{abstract}
\maketitle
\tableofcontents

\section{Geometric construction of exceptional Lie algebras}
\label{sec:geom-constr-except}

The Killing--Cartan classification of complex simple Lie algebras
consists of four infinite series of \emph{classical} Lie algebras:
$A_{n\geq 1}$, $B_{n\geq 2}$, $C_{n\geq 3}$ and $D_{n\geq 4}$, and a
small number of \emph{exceptional} Lie algebras: $E_6$, $E_7$, $E_8$,
$F_4$ and $G_2$, where the subscripts denote the ranks and in the
classical case they have been chosen so as to avoid low rank
isomorphisms.  The classical Lie algebras are well-understood: they
are matrix algebras and their compact real forms are the Lie algebras
of (special) unitary matrices over $\RR$ ($B$ and $D$), $\CC$ ($A$)
and $\HH$ ($C$).  In contrast, the exceptional Lie algebras result
from ``baroque'' constructions involving octonions or else from
constructions involving spinors, as explained by Adams in his
posthumous lecture notes \cite{AdamsExceptional} and, for the case of
$E_8$, also in \cite{GSW}.  It is this latter construction which we
will geometrise in today's lecture, using a device well-known in
supergravity and which will be subject of the next two lectures: the
so-called \emph{Killing superalgebra}.

The basic idea of these lectures is to assign to a spin manifold a
2-graded algebra.   In this first lecture we will consider the
particular example of the exceptional Hopf fibration
\begin{equation}
  \begin{CD}
    S^7 @>>> S^{15} @>>> S^8~,
  \end{CD}
\end{equation}
where $S^n$ stands for the unit $n$-sphere in $\RR^{n+1}$.  If we
think of $S^{15}\subset \OO \oplus \OO$ and $S^7 \subset \OO$, then
$S^8 \cong \OO\PP^1$ is the octonionic projective plane.  Applying the
Killing superalgebra construction to the spaces in the above fibration
we will obtain compact (or split) real forms of the simple Lie
algebras of type $B_4$, $E_8$ and $F_4$, respectively.  We have been
unable thus far to pinpoint the relation between these Lie algebras
which is suggested by the Hopf fibration relating the corresponding
spaces.  This first lecture is based on \cite{JMFE8}.

\subsection{Clifford algebras, spin group and spinor representations}
\label{sec:cliff-algebras}

We start with a flash review of Clifford algebras, the spin group and
the spinor representations.  For more details, see the books \cite{H}
or \cite{LM}.

Let $E, \left<-,-\right>$ be a euclidean vector space.  For instance,
we could take $E=\RR^n$ with the standard ``dot'' product.  We define
the \textbf{Clifford algebra} $\Cl(E)$ to be the associative algebra
obtained by quotienting the tensor algebra $T(E)$ by the two-sided
ideal generated by elements of the form $x \otimes x +
\left<x,x\right> 1$; in symbols,
\begin{equation}
  \Cl(E) = T(E)/\left\{x \otimes x + \left<x,x\right> 1\right\}~.
\end{equation}
Since the ideal is not homogeneous, the Clifford algebra is not
graded, but only filtered.  The associated graded algebra is the
exterior algebra $\Lambda E$, whence we may think of the Clifford
algebra as a quantisation of the exterior algebra.   Since the ideal
has even parity, the Clifford algebra inherits a $2$-grading which
agrees under the ``classical limit'' with the $2$-grading of $\Lambda
E$ into odd and even forms.  When $E=\RR^n$ relative to the usual dot
product, we call the corresponding Clifford algebra $\Cl(n)$.  Up to
isomorphism we have the following table of euclidean Clifford
algebras:
\begin{equation}
  \begin{tabular}[h!]{>{$}r<{$}|*{8}{>{$}c<{$}}|>{$}c<{$}}
    n & 0 & 1 & 2 & 3 & 4 & 5 & 6 & 7 & k+8\\\hline
    \Cl(n) & \RR & \CC & \HH & \HH \oplus \HH & \HH(2) & \CC(4) &
    \RR(8) & \RR(8) \oplus \RR(8) & \Cl(k) \otimes_\RR \RR(16)
  \end{tabular}
\end{equation}
where we use the notation $\KK(m)$ for the algebra of $m\times m$
matrices with entries in $\KK$, and where the last column goes by the
name of \textbf{Bott periodicity}.

The subspace $\Lambda^2E \subset \Cl(E)$ is a Lie subalgebra under the
Clifford commutator isomorphic to $\fso(E)$.  Exponentiating inside
the (associative) Clifford algebra, gives (for $n>2$) a
simply-connected Lie group $\Spin(E) \subset \Cl(E)^{\text{even}}$,
called the \textbf{spin group} of $E$.  Conjugating with $\Spin(E)$
preserves $E \subset \Cl(E)$ and defines a two-to-one group
homomorphism
\begin{equation}
  \Spin(E) \to \SO(E)~.
\end{equation}
When $E=\RR^n$ with the standard dot product, we denote the spin group
by $\Spin(n)$.

The Clifford algebra $\Cl(n)$ is isomorphic either to a matrix algebra
or to two copies of a matrix algebra, and as such has either one or
two inequivalent irreducible representations.  This follows from the
fact that $\RR(n)$ and $\HH(n)$ have up to isomorphism a unique
irreducible representation isomorphic to $\RR^n$ and $\HH^n$,
respectively, whereas $\CC(n)$ has two non-isomorphic irreducible
representations: $\CC^n$ and its complex conjugate representation.
Similarly, $\RR(n) \oplus \RR(n)$ and $\HH(n) \oplus \HH(n)$ have two
inequivalent irreducible representations, isomorphic to $\RR^n$ or
$\HH^n$, respectively.  These are called the \textbf{pinor}
representations of $\Cl(E)$.  Restricting them to $\Spin(E)$ one
obtains (perhaps reducible) representations called \textbf{spinor}
representations and denoted $S(E)$.  The type ($\RR$, $\CC$ or $\HH$)
of $S(E)$ follows from the fact that $\Spin(E) \subset
\Cl(E)^{\text{even}}$ and that $\Cl(n)^{\text{even}} \cong \Cl(n-1)$.
For example, the spinor representations of $\Spin(n)$ for small $n$,
have types $\HH$ for $n=3,4,5$, $\CC$ for $n=2,6$ and $\RR$ for
$n=7,8,9$.  This is consistent with the low-dimensional isomorphisms
$\Spin(2) \cong \U(1)$, $\Spin(3) = \Sp(1)$, $\Spin(4) = \Sp(1) \times
\Sp(1)$, $\Spin(5) = \Sp(2)$, $\Spin(6) = \SU(4)$.

We will be using the fact that $S(E)$ admits a $\Cl(E)$-invariant
inner product $\left(-,-\right)$ obeying
\begin{equation}
  \left(x \cdot \psi_1,\psi_2\right) = - \left(\psi_1, x \cdot
    \psi_2\right)~,
\end{equation}
for all $\psi_i \in S(E)$ and $x \in E \subset \Cl(E)$.  This means
that $\left(-,-\right)$ is also $\Spin(E)$-invariant.  For $E$
euclidean, as we have been assuming, the spinor inner product is also
positive-definite.  However if we allow for $E$ to have arbitrary
signature, then all seven types of elementary inner products appear
among the spinor inner products.

The transpose of the Clifford action $E \otimes S(E) \to S(E)$ defines
a real bilinear map $S(E) \otimes S(E) \to E$ which in the cases of
interest in this lecture will be skewsymmetric, whence defines a map
$\Lambda^2 S(E) \to E$.  This will form part of a Lie bracket on a
2-graded algebra whose odd subspace will be isomorphic to (a subspace
of) $S(E)$.

\subsection{Globalisation: spin geometry}
\label{sec:glob-spin-geom}

Let $(M^n,g)$ be a riemannian manifold.  At every point $x \in M$ we
can consider the orthonormal frames for the tangent space $T_xM$.
This is the fibre at $x$ of a principal fibre bundle $\O(M)$ called,
unsurprisingly, the bundle of orthonormal frames.  If $M$ is oriented,
and restricting to \emph{oriented} frames, we obtain a subbundle
$\SO(M)$.  The obstruction to the existence of $\SO(M)$ is the
triviality of $\det(TM)$ which is captured by the first
Stiefel--Whitney class $w_1(TM)\in H^1(M;\ZZ_2)$.  Hence roughly
speaking half the manifolds are orientable.  Assuming that $M$ is
oriented, a spin structure on $M$ is a lift $\Spin(M) \to \SO(M)$
restricting fibrewise to the two-to-one homomorphism $\Spin(n) \to
\SO(n)$.  The obstruction is now captured by the second
Stiefel--Whitney class $w_2(TM) \in H^2(M;\ZZ_2)$, whence roughly
speaking one quarter of all manifolds are spin.  If $w_2(TM) = 0$,
the set of inequivalent spin structures are in one-to-one
correspondence with $H^1(M;\ZZ_2) \cong \Hom(\pi_1(M),\ZZ_2)$, so
roughly speaking to the assignment of a sign to every noncontractible
loop.

For example, if $S^n \subset \RR^{n+1}$ is the unit sphere, then $T_x
S^n$ is the perpendicular complement of the line in $\RR^{n+1}$
through the origin and $x$.  An oriented orthonormal basis for $T_x
S^n$ is then an oriented orthonormal frame for $x^\perp \cong \RR^n$
and hence in one-to-one correspondence with the points in $\SO(n)$.
Adding $x$ itself we obtain an oriented orthonormal frame for
$\RR^{n+1}$ and hence an element of $\SO(n+1)$: the element which
takes the standard orthonormal basis to that one.  Conversely, we have
a map $\SO(n+1) \to S^n$ sending the matrix $g \in \SO(n+1)$ to its
first column, say $x$, which is a unit vector in $\RR^{n+1}$.  The
fibre of this map consists of the remaining $n$ columns, which form an
oriented frame in the $n$-dimensional subspace perpendicular to $x$.
In other words, $\SO(S^n) = \SO(n+1)$.  The spin cover $\Spin(S^n)$ is
precisely the spin group $\Spin(n+1)$ and since $\pi_1(S^n) =
\left\{1\right\}$ (for $n>1$), there is a unique such spin structure.
For $n=1$ there are two spin structures, which physicists like to call
Neveu--Schwarz and Ramond \cite{GSW}.

Let $\rho: \Spin(n) \to \GL(S)$ be a spinor representation of
$\Spin(n)$ and define the associated vector bundle
\begin{equation}
  \$ = \Spin(M) \times_\rho S~,
\end{equation}
called a bundle of spinors.  Its sections are called spinor fields.
(Had the lectures been given in Spanish, the spinor bundle would have
been called €!)

The tangent space $(T_xM, g_x)$ to $M$ at $x$ is a euclidean vector
space and gives rise to a Clifford algebra $\Cl(T_xM)$.  As $x$
varies, this globalises to a bundle $\Cl(TM)$ of Clifford algebras.
As a vector bundle, we have a natural isomorphism $\Cl(TM) \cong
\Lambda T^*M$.  We will always think of spinor representations as the
restriction to the spin group of a pinor representation of the
Clifford algebra.  Globalising, this means that the spinor bundle $\$$
will always admit an action of the Clifford bundle $\Cl(TM)$, making
it into a bundle of Clifford modules.  In this way, differential
forms, which are sections of $\Lambda T^*M$, will be able to act on
spinor fields via the natural isomorphism $\Lambda T^*M \cong
\Cl(TM)$ and the action of $\Cl(TM)$ on $\$$.

The Levi-Cività connection on $\SO(M)$ lifts to a connection on
$\Spin(M)$ and hence defines a connection on any associated vector
bundle.  In particular we have a covariant derivative $\nabla$ on
sections of $\$$: for all vector fields $X\in\eX(M)$ and spinor fields
$\psi \in \Gamma(\$)$, $\nabla_X\psi \in \Gamma(\$)$.  The covariant
derivative $\nabla_X$ along $X$ is linear and obeys the Leibniz rule
\begin{equation}
  \nabla_X(f\psi) = (X f) \psi + f \nabla_X \psi~,
\end{equation}
for all functions $f \in C^\infty(M)$.

\subsection{Killing spinors and the cone construction}
\label{sec:killing-spinors-cone}

The Levi-Cività connection may be used to write down natural equations on
spinors, whose solutions define privileged notions of spinor fields:
\begin{itemize}
\item \textbf{parallel spinors}: $\nabla \psi = 0$.  By the holonomy
  principle, the holonomy group of $\nabla$ must be included in the
  stabilizer of a spinor.  The determination of which manifolds admit
  parallel spinors was thus solved by Wang \cite{Wang} using Berger's
  holonomy classification.   The irreducible holonomy groups of
  manifolds admitting parallel spinors are $\SU(n)$, $\Sp(n)$, $G_2$
  and $\Spin(7)$.
\item \textbf{Killing spinors}: $\nabla_X \psi = \lambda X \cdot \psi$
  for all $X \in \eX(M)$ for some nonzero constant $\lambda \in \CC$,
  called the \textbf{Killing constant}.  Iterating the definition of a
  Killing spinor, we find that $M$ is Einstein with scalar curvature
  proportional to $\lambda^2$, which means that $\lambda^2 \in \RR$
  and hence $\lambda \in \RR^\times \cup i \RR^\times$.  This gives
  rise to two separate notions of \textbf{real} or \textbf{imaginary}
  Killing spinors, according to whether $\lambda$ is real or
  imaginary, respectively.
\end{itemize}

In this lecture we will concentrate on the case of real Killing
spinors.  Moreover by rescaling the metric, if necessary, we may
always take $\lambda = \pm\half$.  Therefore such a manifold $M$ is
Einstein with positive scalar curvature and, if complete, is compact
by the Bonnet--Myers theorem.  The question of which complete spin
manifolds admit real Killing spinors was solved by Bär \cite{Baer} via
his celebrated cone construction by mapping the problem to the problem
of determining which manifolds admit parallel spinors.

Indeed, given a spin manifold $(M,g)$ we define its \textbf{metric
  cone} $C(M) = \RR^+ \times M$, with metric
\begin{equation}
  g_C = dr^2 + r^2 g~,
\end{equation}
where $r>0$ is the parameter of the $\RR^+$. For example, if $M=S^n$,
then $C(M) = \RR^{n+1}\setminus\{0\}$.  In this case, and in this case
alone, the metric extends smoothly to the origin and it is the flat
metric on $\RR^{n+1}$ written in spherical polar coordinates.  In all
other cases, the metric has a conical singularity at $r=0$.  Bär's
penetrating observation was that $\nabla_X \psi = \pm \half X \cdot
\psi$ on $M$ becomes the condition $\widetilde\nabla \widetilde\psi =
0$ on the cone, where the tilded objects live on the cone.  The sign
in the Killing spinor equation is reflected either in the chirality of
the parallel spinor in the case of $n$ odd, or in the embedding
$\Cl(n) \subset \Cl(n+1)$ if $n$ is even.  This says that the
existence of real Killing spinors is again a holonomy problem, albeit
in an auxiliary manifold one dimension higher.  A theorem of Gallot
\cite{Gallot} says that if $M$ is complete, then its metric cone is
either flat --- so that $M$ is the round sphere --- or irreducible.
In this latter case, we may use Wang's classification of holonomy
groups leaving a spinor invariant.  In this way one arrives at Table
\ref{tab:cone} of (types of) complete manifolds admitting real Killing
spinors.

\begin{table}[h!]
  \centering
  \caption{Manifolds admitting real Killing spinors}
  \begin{tabular}[h!]{>{$}c<{$}|>{$}c<{$}|>{$}c<{$}|>{$}c<{$}}
    n & \text{hol}~\widetilde\nabla & C(M) & M\\ \hline
    n & \left\{1\right\} & \RR^{n+1} & S^n \\
    2m - 1 & \SU(m) & \text{Calabi--Yau} & \text{Sasaki--Einstein}\\
    4m - 1 & \Sp(m) & \text{hyperkähler} & \text{3-Sasaki}\\
    6 & G_2 &  & \text{nearly Kähler (non-Kähler)}\\
    7 & \Spin(7) &  &  \text{weak $G_2$ holonomy}
  \end{tabular}
  \label{tab:cone}
\end{table}

Such manifolds play an important rôle in the AdS/CFT correspondence,
as pointed out originally in \cite{AFHS}.

\subsection{The Killing superalgebra}
\label{sec:kill-super-algebra}

Let $(M,g)$ be a spin manifold and $\$ \to M$ a bundle of
$\Cl(TM)$-modules.  We will assume that $M$ admits real Killing
spinors and, without loss of generality (i.e., rescaling the metric
and reversing orientation, if necessary), assume that the Killing
constant $\lambda = \half$.  We define a $2$-graded vector space $\fg
= \fg_0 \oplus \fg_1$, where $\fg_0$ is the vector space of Killing
vector fields on $M$ and
\begin{equation}
  \fg_1 = \left\{\psi \in \Gamma(\$) \middle | \nabla_X \psi = \half X
  \cdot \psi~\forall\, X \in \eX(M)\right\}
\end{equation}
is the vector space of Killing spinors.  Remember we have a real
bilinear map $\fg_1 \times \fg_1 \to \eX(M)$ by transposing the
Clifford action.  Explicitly, given $\psi_1,\psi_2 \in \fg_1$, we let
$[\psi_1,\psi_2] \in \eX(M)$ be defined by
\begin{equation}
  g([\psi_1,\psi_2], X) = \left(\psi_1, X \cdot \psi_2\right)
\end{equation}
for all $X \in \eX(M)$.  The following result explains the
terminology.

\begin{lemma}
  $[\psi_1,\psi_2]$ is a Killing vector field.
\end{lemma}

\begin{proof}
  This is a simple calculation.  For all $X,Y\in\eX(M)$, we have
  \begin{align*}
    g(\nabla_X [\psi_1,\psi_2], Y) &= X\, g([\psi_1,\psi_2], Y)
    -  g([\psi_1,\psi_2], \nabla_X Y)\\
    &= X\, \left(\psi_1, Y \cdot \psi_2\right) - 
    \left(\psi_1, \nabla_X Y \cdot \psi_2\right)\\
    &= \left(\nabla_X \psi_1, Y \cdot \psi_2\right) +
    \left(\psi_1,  Y \cdot \nabla_X\psi_2\right)\\
    &= \half \left(X \cdot \psi_1, Y \cdot \psi_2\right) + 
    \half \left(\psi_1,  Y \cdot X\cdot \psi_2\right)\\
    &= \half \left(\psi_1, (Y \cdot X - X \cdot Y)  \cdot
      \psi_2\right)~,
  \end{align*}
  whence
  \begin{equation}
    g(\nabla_X [\psi_1,\psi_2], Y)  + g(\nabla_Y [\psi_1,\psi_2],X)  = 0~.
  \end{equation}
\end{proof}

Let $K \in \fg_0$ be a Killing vector field.  Then $A_K : TM \to TM$,
defined by $A_K(Y) = - \nabla_Y K$, is a skewsymmetric endomorphism of
$TM$.  Define the following \textbf{Lie derivative} on spinor fields:
\begin{equation}\label{eq:Lieder}
  \eL_K = \nabla_K + \varrho(A_K)~,
\end{equation}
where $\varrho: \fso(TM) \to \fgl(\$)$ is the spinor representation at
the level of the Lie algebra.

Taking for $\varrho$ any other representation of $\fso(TM)$ defines
equally well a Lie derivative on sections of the corresponding
associated vector bundle.  For example, if we take $\varrho$ to be the
defining representation on $TM$, we have
\begin{equation}
  \eL_K X = \nabla_K X + A_K X = \nabla_K X - \nabla_X K = [K,X]~,
\end{equation}
which is the standard Lie derivative of vector fields.

\begin{proposition}\label{pr:Lieder}
  The Lie derivative $\eL_K$ on $\Gamma(\$)$ obeys a number of
  properties:
  \begin{enumerate}
  \item $[\eL_{K_1},\eL_{K_2}] \psi = \eL_{[K_1,K_2]}\psi$,
  \item $\eL_K (f \psi) = (K\, f) \psi + f \eL_K \psi$,
  \item $\eL_K (X \cdot \psi) = [K,X] \cdot \psi + X \cdot \eL_K
    \psi$, and
  \item $\eL_K \nabla_X \psi = \nabla_X \eL_K \psi + \nabla_{[K,X]} \psi$~,
  \end{enumerate}
  for all Killing vector fields $K,K_1,K_2$ and all $f \in
  C^\infty(M)$, $X\in \eX(M)$ and $\psi \in \Gamma(\$)$.
\end{proposition}

We remark that properties (1) and (2) justify calling $\eL_K$ a Lie
derivative, whereas (3) and (4) say that $\eL_X$ leaves invariant the
Clifford action and $\nabla$, respectively.  The definition of $\eL_K$
goes back to Lichnerowicz and appears in the thesis of
Kosmann-Schwarzbach \cite{Kosmann}.  It appeared also more recently,
in a supergravity context, in \cite{JMFKilling}.

The last two properties in Proposition \ref{pr:Lieder} imply that if
$\psi\in\fg_1$ is a Killing spinor, then so is $\eL_K\psi$ for all
Killing vector fields $K$.  Indeed,
\begin{align*}
  \nabla_X \eL_K \psi &= \eL_K \nabla_X \psi - \nabla_{[K,X]}\psi &&
  \text{by (4) in Proposition \ref{pr:Lieder}}\\
  &= \half \eL_K ( X \cdot \psi) - \half [K,X]\cdot \psi && \text{since
    $\psi \in \fg_1$}\\
  &= \half X \cdot \eL_K && \text{by (3) in Proposition \ref{pr:Lieder}}
\end{align*}
whence $\eL_K \psi \in \fg_1$.  This defines a real bilinear map
$\fg_0 \times \fg_1 \to \fg_1$, denoted $(K,\psi) \mapsto [K,\psi] :=
\eL_K \psi$.

We now have defined a 2-graded multiplication on $\fg = \fg_0 \oplus
\fg_1$, denoted by a bracket $[-,-]$ anticipating the fact that in
some cases it will be a Lie (super)algebra.  To wit, we have a map
$\Lambda^2 \fg_0 \to \fg_0$, given by the Lie bracket of vector
fields, the above-defined map $\fg_0 \otimes \fg_1 \to \fg_1$ given by
the spinorial Lie derivative, and the map $\fg_1 \otimes \fg_1 \to
\fg_0$ given by transposing the Clifford action.  This last map is
either symmetric or skewsymmetric depending on dimension.  This
suggests that it may define a Lie (super)algebra structure on $\fg$.
This requires satisfying the relevant Jacobi identity.  Being a
2-graded algebra, the \emph{jacobator}, the element in
$\Hom(\fg^{\otimes 3}, \fg)$ whose vanishing implies the Jacobi
identity, breaks up into four components depending on whether we have
three, two, one or no elements in $\fg_0$:
\begin{itemize}
\item all elements in $\fg_0$: this is simply the Jacobi identity of
  the Lie bracket of vector fields;
\item two elements in $\fg_0$ and one in $\fg_1$:
  \begin{align*}
    [K_1,[K_2,\psi]] - [K_2,[K_1,\psi]] &= \eL_{K_1} \eL_{K_2} \psi -
    \eL_{K_2} \eL_{K_1} \psi  \\
    &= [\eL_{K_1},\eL_{K_2}] \psi\\
    &= \eL_{[K_1,K_2]} \psi && \text{by (1) in Proposition
      \ref{pr:Lieder}}\\
    &= [[K_1,K_2],\psi]~;
  \end{align*}
\item one element in $\fg_0$ and two in $\fg_1$: this is property (3)
  in Proposition \ref{pr:Lieder}; and
\item all elements in $\fg_1$: this does \emph{not} follow from the
  formalism and has to be checked case by case.  For Lie algebras, it
  lives in $\left(\Lambda^3\fg_1^* \otimes \fg_1\right)^{\fg_0}$,
  whereas for Lie superalgebras it lives in $\left(S^3\fg_1^* \otimes
    \fg_1\right)^{\fg_0}$.  In some cases, representation theory shows
  that such spaces are $0$, and hence this last component of the
  jacobator vanishes.  In other cases, such spaces are not $0$, but
  the jacobator vanishes all the same.  In most cases, however, this
  last component of the jacobator will not vanish.  Hence the generic
  situation is a 2-graded \textbf{$\boldsymbol{\frac{3}{4}}$-Lie
    (super)algebra}.
\end{itemize}

\subsection{Some examples of 2-graded Lie algebras}
\label{sec:some-examples-2}

Consider now the unit spheres $S^7 \subset \RR^8$, $S^8 \subset \RR^9$
and $S^{15} \subset \RR^{16}$, thought of as riemannian
manifolds with the canonical spin structure given by the spin groups.
In all these cases, the spinor inner product is real symmetric and
positive-definite.  Since Clifford action is skewsymmetric, so is its
transpose, whence the odd-odd bracket is similarly skewsymmetric,
defining a map $\Lambda^2 \fg_1 \to \fg_0$.  In other words, $\fg$ is a
2-graded (possibly) Lie algebra.  Notice that $[\fg_1,\fg_1]$ is an
ideal of $\fg_0$: this does not use the vanishing of the last
component of the jacobator.  Since $\fg_0$, the isometry Lie algebra
of the above spheres, is simple, we see that $[\fg_1,\fg_1] = \fg_0$
in this case.

To determine $\fg_1$ as an $\fg_0$-module we use the cone construction
and the fact, proved in \cite{JMFKilling}, that this construction is
equivariant under the action of $\fg_0$, which is naturally a Lie
subalgebra of the isometries of the cone.  This means that $[K,\psi] =
\eL_K \psi$ can be lifted and calculated on the cone:
\begin{equation}
  \eL_{\widetilde K} \widetilde\psi = \widetilde\nabla_{\widetilde K}
  \widetilde\psi + \widetilde\varrho(A_{\widetilde K}) \widetilde\psi~,
\end{equation}
but $\widetilde\psi$ is parallel and (relative to flat coordinates on
the cone) $\widetilde\psi$ and $A_{\widetilde K}$ are constant because
$\widetilde K$ is a linear vector field.  Therefore this is the
standard action of $\fg_0 = \fso(n+1)$ ($n=7,8,15$) on
(positive-chirality, when applicable) spinors: $S(8)_+$, $S(9)$ and
$S(15)_+$, all of which are irreducible representations.

In all cases, a roots-and-weights calculation (made less painful by
using LiE \cite{LiE}) shows that
\begin{equation}
  \left(\Lambda^3\fg_1^* \otimes \fg_1\right)^{\fg_0} = 0~,
\end{equation}
whence the Jacobi identity is satisfied and $\fg$ becomes a 2-graded
Lie algebra.  To identify the Lie algebras in question we simply
observe that $\fg_0$ being simple and $\fg_1$ being irreducible,
implies that $\fg$ is simple.  The dimensions are easy to compute and
the Lie algebras are thus easy to recognise from the Killing--Cartan
classification.  The results are summarised in Table
\ref{tab:KSAspheres}.

\begin{table}[h!]
  \centering
  \caption{Killing Lie algebras of some spheres}
  \begin{tabular}{*{6}{>{$}c<{$}|}>{$}c<{$}}
    M & \fg_0 & \dim\fg_0 & \fg_1 & \dim\fg_1 & \dim\fg & \fg\\ \hline
    S^7 & \fso(8) & 28 & S(8)_+ & 8 & 36 & B_4\\
    S^8 & \fso(9) & 36 & S(9) & 16 & 52 & F_4\\
    S^{15} & \fso(16) & 120 & S(16)_+ & 128 & 248 & E_8
  \end{tabular}
  \label{tab:KSAspheres}
\end{table}

Since the inner products on $\fg_0$ and $\fg_1$ are invariant and
positive-definite, $\fg$ is a compact real form of the corresponding
complex simple Lie algebra.  By the usual device of taking $i\fg_1$
instead of $\fg_1$, we may obtain the maximally split real forms.

\section{Supergravity backgrounds}
\label{sec:supergr-backgr}

Supergravity is an extension of Einstein (or Einstein--Maxwell)
theory.  At the level of its solutions, it is given by some geometric
data $(g,F,\dots)$, where $g$ is a local lorentzian metric and
$F,\dots$ stand for extra fields, all subject to partial differential
equations of the form
\begin{itemize}
\item \emph{Einstein}
  \begin{equation}
    \Ric(g) - \half R g = T(F,\dots)
  \end{equation}
\item \emph{``Maxwell''}
  \begin{equation}
    d F = 0 \qquad\text{and}\qquad d \star F = \cdots
  \end{equation}
\end{itemize}
The details depend on the supergravity theory in question and have
hence kept purposefully vague in the above description.  We will
consider here only so-called \emph{Poincaré} supergravities.  There
are other supergravities: massive, gauged,...  Supergravity theories
are dictated by the representation theory of the Poincaré
superalgebras.  There are (physically interesting) supergravity
theories in dimension $d\leq 11$ and lorentzian signature, meaning
that the local metric $g$ is lorentzian.  Supergravity theories are
among the jewels of twentieth century theoretical physics and a good
review of the structure of supergravity theories from the
representation theory point of view can be found in
\cite{ToineReview}.

\subsection{Eleven-dimensional supergravity}
\label{sec:elev-dimens-supergr}

My favourite, and to some extent the simplest yet nontrivial,
supergravity theory is the unique eleven-dimensional supergravity
theory.  Its existence was conjectured by Nahm \cite{Nahm}, whereas it
was constructed by Cremmer, Julia and Scherk \cite{CJS}.  Its field
content is a lorentzian eleven-dimensional metric $g$ and a closed
4-form $F$.  We can motivate this as follows.

Supergravity is a theory invariant under local supersymmetry, hence
the spectrum should carry a representation of the corresponding
supersymmetry algebra.  In the case of eleven-dimensional supergravity
this is the eleven-dimensional Poincaré superalgebra $(\fso(1,10)
\oplus \RR^{1,10}) \oplus S(1,10)$, where $S(1,10)$ is the spinor
representation of $\Spin(1,10)$.  It is not hard to show, using Bott
periodicity, that $\Cl(1,10) = \RR(32) \oplus \RR(32)$ and hence
$S(1,10) \cong \RR^{32}$.  The supertranslation ideal generated by
$\RR^{1,10} \oplus S(1,10)$ has as nonzero brackets the projection
$S^2 S(1,10) \to \RR^{1,10}$ of the symmetric square of the spinor
representation $S^2 S(1,10) \cong \RR^{1,10} \oplus
\Lambda^2\RR^{1,10} \oplus \Lambda^5\RR^{1,10}$ onto the vector
representation of $\Spin(1,10)$.

Irreducible unitary representations of the Poincaré superalgebra are
induced by representations of the supertranslation ideal generated by
$S(1,10) \oplus \RR^{1,10}$.  This is done by first fixing a character
of the abelian translation ideal $\RR^{1,10}$; that is, a momentum $p
\in \left(\RR^{1,10}\right)^*$.  Since we are interested in a theory
of gravity, which we expect even in eleven dimensions to be a long
range force, we require a massless representations, whence $p^2 = 0$,
but $p \neq 0$.  The \emph{little group} of $p$, which is the maximal
compact subgroup of the stabiliser of $p$ in $\Spin(1,10)$ is
isomorphic to $\Spin(9)$.  Once a momentum $p$ has been fixed, the
supertranslation ideal takes the form of a Clifford algebra
\begin{equation}\label{eq:cliffp}
  [Q_1,Q_2] = - 2 \left(Q_1, p \cdot Q_2\right) 1~,
\end{equation}
where $Q_i \in S(1,10)$ and the spinor inner product is symplectic in
this signature, whence the bracket here is symmetric.  The bilinear
form defining the Clifford algebra, $\left<Q_1,Q_2\right> = \left(Q_1,
  p \cdot Q_2\right)$ is degenerate because $p^2 = 0$.  In fact, it
has rank $16$.  This is shown by exhibiting a ``dual'' momentum $q$
such that $q^2 = 0$ and $p \cdot q = 1$.  Then $S(1,10) = \ker p
\oplus \ker q$, where the kernel refers to the Clifford action.  The
common stabiliser of $p$ and $q$ in $\Spin(1,10)$ is a
$\Spin(9)$-subgroup, which we can identify with the little group of
either $p$ or $q$.  The degenerate Clifford algebra \eqref{eq:cliffp}
becomes an honest Clifford algebra on $\ker q \subset S(1,10)$
isomorphic to $\Cl(16)$, and in fact, as $\Spin(9)$-module, $\ker q$
is the spinor module $S(9)$.  There is, up to isomorphism, a unique
irreducible representation of $\Cl(16)$ and it is real and of
dimension $256$.  Indeed, by Bott periodicity,
\begin{equation}
  \Cl(16) \cong \Cl(8) \otimes_\RR \RR(16) \cong \RR(16) \otimes_\RR
  \RR(16) \cong \RR(16^2)~.
\end{equation}
As a $\Spin(9)$-module, this is nothing but $S(S(9))$; that is,
spinors of spinors!  A roots-and-weights calculation shows that as a
representation of $\Spin(9)$ we have
\begin{equation}
  S(S(9)) \cong S^2_0(\RR^9) \oplus \Lambda^3 \RR^9 \oplus
  \text{RS}(\RR^9)~,
\end{equation}
where $\RR^9$ stands for the vector representation of $\Spin(9)$,
$S^2_0$ denotes traceless symmetric tensors and $\text{RS}$ stands for
the Rarita--Schwinger representation, which is the subrepresentation
of $\RR^9 \otimes S(9)$ consisting of the kernel of the Clifford
action $\RR^9 \otimes S(9) \to S(9)$.  Counting dimensions, we see that
for the bosonic part of the representation
\begin{equation}
  \dim S^2_0(\RR^9) + \dim \Lambda^3 \RR^9  = 44 + 84 = 128~,
\end{equation}
whereas for the fermionic part of the representation
\begin{equation}
  \dim \text{RS}(\RR^9) = \dim \RR^9 \otimes S(9) - \dim S(9) = 16
  \times 9 - 16 = 128~,
\end{equation}
whence the physical degrees of freedom match, as expected.  In terms
of fields, $S^2_0(\RR^9)$ parametrise the fluctuations of a metric
tensor $g$, whereas $\Lambda^3 \RR^9$ parametrises the fluctuations of
a (locally defined) 3-form potential $A$, and $\text{RS}(\RR^9)$
parametrises the fluctuations of a gravitino.

The supergravity action, ignoring terms involving the gravitino,
consists of three terms: an Einstein--Hilbert term, a Maxwell term and
a Chern--Simons term.  The lagrangian density is given by
\begin{equation}\label{eq:action}
  R \dvol_g - \tfrac14 F \wedge \star F + \tfrac1{12} F \wedge F
  \wedge A~,
\end{equation}
where $F = dA$ locally.  Although $A$ appears explicitly in the above
lagrangian, the Euler--Lagrange equations only involve $F$.  The
equations are of Einstein--Maxwell type with a twist provided by the
Chern--Simons term; namely, the Maxwell equation is nonlinear:
\begin{equation}
  d \star F = -\half F \wedge F~.
\end{equation}

\subsection{Supersymmetric supergravity backgrounds}
\label{sec:supersymm-supergr-ba}

We define a (bosonic) \textbf{eleven-dimensional supergravity
  background} to be an eleven-dimensional lorentzian spin manifold
$(M,g,\$)$ and a closed 4-form $F \in \Omega^4(M)$ subject to the
Einstein--Maxwell equations derived from the lagrangian
\eqref{eq:action}.

The lagrangian \eqref{eq:action} admits a supersymmetric completion by
adding extra terms involving the gravitino $\Psi \in \Omega^1(M,\$)$.
The variation of the gravitino under supersymmetry defines a
connection $D$ on the spinor bundle $\$$:
\begin{equation}
  D_X \psi := \nabla_X \psi + \tfrac16 \iota_X F \cdot \psi +
  \tfrac1{12} X^\flat \wedge F \cdot \psi~,
\end{equation}
for all $\psi \in \Gamma(\$)$ and $X \in \eX(M)$ and where $X^\flat
\in \Omega^1(M)$ is the one-form such that $X^\flat(Y) = g(X,Y)$ for
all $Y \in \eX(M)$.

The connection $D$ is the fundamental object in this game, as it encodes
virtually all the information of the theory.  For example, the
Einstein--Maxwell equations can be recovered by demanding the
vanishing of the Clifford-trace of its curvature.  More explicitly,
let $e_i$ be a pseudo-orthonormal frame for $M$ and let $e^i$ denote
the dual frame, defined by $g(e^i,e_j) = \delta^i_j$.  Then, as shown
in \cite{GauPak}, the field equations defining the notion of a
supergravity background are equivalent to
\begin{equation}
  \sum_i e^i \cdot R^D(e_i,X) = 0 \quad \forall~X \in \eX(M)~.
\end{equation}

A nonzero spinor field  $\psi \in \Gamma(\$)$ which is $D$-parallel
is called a \textbf{(supergravity) Killing spinor}.  Although this
seems a priori to be a generalisation of the notion of a parallel
spinor, it is in fact the original notion of a Killing spinor.  The
geometrical notion in the first lecture is a special case of the
supergravity Killing spinor equation for a particular Ansatz for
$(M,g,F)$, known as a \emph{Freund--Rubin background}
\cite{FreundRubin}.

Being a linear equation, Killing spinors form a vector space, which
anticipating the construction of the Killing superalgebra, will be
denoted $\fg_1$.  Being defined by a parallel condition, a Killing
spinor is determined by its value at a point, whence the dimension of
$\fg_1$ is bounded above by the rank of the spinor bundle; that is,
$\dim\fg_1 \leq 32$.  The ratio
\begin{equation}
  \nu = \frac{\dim \fg_1}{32}
\end{equation}
is called the \textbf{supersymmetry fraction} of the background
$(M,g,F)$.  If $\nu >0$, $(M,g,F)$ is said to be
\textbf{supersymmetric}.

\subsection{Examples}
\label{sec:examples}

A large number of supersymmetric backgrounds are known.  Maximally
supersymmetric backgrounds -- those with $\nu =1$ --- have been
classified in \cite{FOPflux,Bonn,FOPMax}.  For such backgrounds, $D$
is flat and the equations of motions are automatically satisfied.
These backgrounds are related as follows:
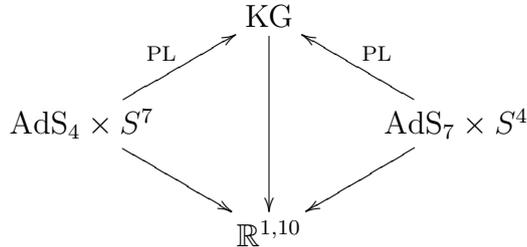
\begin{equation*}
  \xymatrix{
   &  \text{KG} \ar[dd] & \\
    \text{AdS}_4 \times S^7 \ar[ur]^{\text{PL}} \ar[dr] & &
    \text{AdS}_7 \times S^4 \ar[ul]_{\text{PL}} \ar[dl]\\
   & \RR^{1,10} &
  }
\end{equation*}
where $\text{AdS}_n$ is the $n$-dimensional anti~de~Sitter spacetime
--- i.e., lorentzian hyperbolic space ---, KG is a special type of
plane wave \cite{KG} whose geometry is described by a lorentzian
symmetric space of Cahen--Wallach type \cite{CahenWallach}, and
$\RR^{1,10}$ is Minkowski spacetime with $F=0$.  The arrows labelled
``PL'' are Penrose limits, described in this context in
\cite{ShortLimits,Limits}, but tracing their origin to work of Güven
\cite{GuevenPlaneWave} and, of course, Penrose
\cite{PenrosePlaneWave}.  The undecorated arrows are zero-curvature
limits.

The $\text{AdS}_4 \times S^7$ and $\text{AdS}_7 \times S^4$
backgrounds depend on a parameter, interpreted as the scalar curvature
of the eleven-dimensional geometry.  The radii of curvature of the
factors are in a ratio of $2:1$, whence these backgrounds do not
describe realistic compactifications as they once were thought to do.
They are known as Freund--Rubin backgrounds.  The Killing spinors of
the Freund--Rubin backgrounds are $\otimes$ of geometric Killing
spinors on the two factors: real on the riemannian factor and
imaginary on the lorentzian factor.  One can substitute either factor
by an Einstein manifold with the same scalar curvature and admitting
the relevant kind of Killing spinors.  In particular one can consider
$\text{AdS}_4 \times X^7$, where $X$ is a riemannian manifold
admitting real Killing spinors, whence its cone has holonomy contained
in $\Spin(7)$.  Whenever $X$ is not a sphere, the resulting background
has a smaller fraction $\nu$ of supersymmetry.  They can be understood
as \emph{near-horizon} geometries of M2-branes, to which we now turn.

The \textbf{M2-brane} is a interesting background with $\nu=\half$,
discovered in \cite{DS2brane} and interpreted as an
\emph{interpolating soliton} in \cite{DuffGibbonsTownsend}.  It is
described as follows:
\begin{equation}
  \begin{aligned}[m]
    g &= H^{-2/3} ds^2(\RR^{1,2}) + H^{1/3} \left(dr^2 + r^2
      ds^2(S^7)\right)\\
    F &= \dvol(\RR^{1,2}) \wedge dH^{-1}\\
    H(r) &= \alpha + \frac{\beta}{r^6}~,
  \end{aligned}
\end{equation}
where $ds^2(\RR^{1,2})$ and $\dvol(\RR^{1,2})$ are the metric and
volume of 3-dimensional Minkowski spacetime, $ds^2(S^7)$ is the metric
on the unit sphere in $\RR^8$ and $H$ is a two-parameter harmonic
function on $\RR^8$.  If we take $\beta\to 0$ while keeping $\alpha
\neq 0$ fixed, we obtain eleven-dimensional Minkowski spacetime with
$F=0$, but taking $\alpha \to 0$ while keeping $\beta \neq 0$ fixed,
one obtains $\text{AdS}_4 \times S^7$ with scalar curvature depending
on $\beta$.  Therefore the M2-brane interpolates between these two
maximally supersymmetric backgrounds.   The Killing spinors are given by
\begin{equation}
  \psi = H^{1/6} \psi_\infty~,
\end{equation}
where $\psi_\infty$ is a parallel spinor in the asymptotic Minkowski
spacetime obeying the projection condition
\begin{equation}
  \dvol(\RR^{1,2}) \cdot \psi_\infty = \psi_\infty~.
\end{equation}
Since $\dvol(\RR^{1,2})^2 = 1$ and the parallel spinors of
$\RR^{1,10}$ split into two half-dimensional eigenspaces of
$\dvol(\RR^{1,2})$.  As a result the solution has $\nu = \half$.

As observed in \cite{AFHS}, replacing $S^7$ by another
seven-dimensional manifold admitting real Killing spinors --- that is,
weak $G_2$-holonomy, Sasaki-Einstein or 3-Sasaki manifolds --- we
obtain an M2-brane at a conical singularity in an 8-dimensional
manifold with $\Spin(7)$, $\SU(4)$ or $\Sp(2)$ holonomy, respectively.

To this day a large class class of backgrounds with various values of
$\nu$ are known to exist.  General local metrics with minimal
supersymmetry have been written down in \cite{GauPak,GauGutPak}.  To
date, the only fraction which has been ruled out is $\nu =
\frac{31}{32}$ \cite{NoMPreons,FigGadPreons}.

\section{The Killing superalgebra of supergravity backgrounds}
\label{sec:kill-super-supergr}

In the first lecture we saw that from a spin manifold admitting
Killing spinors one could define (in the good cases) a 2-graded Lie
algebra and in this way we recovered the compact real forms of the
simple Lie algebras of types $B_4$, $F_4$ and $E_8$.  At its most
basic, what we have is a spin manifold with a privileged subspace of
spinor fields which then generates a 2-graded algebra with the spinors
being the odd-subspace.

In the second lecture we saw how supersymmetric (eleven-dimensional)
supergravity backgrounds gave rise to precisely such a situation: an
eleven-dimensional \emph{lorentzian} spin manifold with a privileged
notion of spinor: the supergravity Killing spinors, which are parallel
with respect to a connection $D$ on the spinor bundle.  Unlike the
spin connection $\nabla$, the connection $D$ is \emph{not} induced
from a connection on the tangent bundle: it is genuinely a spinor
connection.  The spinor bundle is a real rank-32 symplectic vector
bundle, but $D$ does \emph{not} preserve the symplectic structure.  In
fact, as shown by Hull \cite{HullHolonomy}, the holonomy algebra of
$D$ is generically contained in $\fsl(32,\RR)$ since only the
`determinant' is preserved.

In this third and last lecture we will see that to every
supersymmetric background of eleven-dimensional supergravity one can
assign a Lie superalgebra by the techniques in the first lecture and
using it we will show that if $\nu$ is sufficiently large, the
background is forced to be homogeneous.  This lecture is based on
\cite{FMPHom}.

\subsection{The Killing superalgebra}
\label{sec:killing-superalgebra}

Let $(M,g,F)$ be a supersymmetric eleven-dimensional supergravity
background.  Following the method in the first lecture, let us define
a 2-graded vector space $\fg = \fg_0 \oplus \fg_1$, where
\begin{equation}
  \fg_0 = \left\{ X \in \eX(M) \middle | \eL_X g = 0 = \eL_X F
  \right\}
\end{equation}
is the Lie algebra of $F$-preserving isometries of the background, and
\begin{equation}
  \fg_1 = \left\{\psi \in \Gamma(\$) \middle | D \psi = 0\right\}
\end{equation}
is the space of Killing spinors.   Clearly $\fg$ is
finite-dimensional, since as mentioned above $\dim \fg_1 \leq 32$ and
$\dim \fg_0 \leq 66$, which is the maximum dimension of the isometry
algebra of an eleven-dimensional lorentzian manifold.  It is only for
$\RR^{1,10}$ with $F=0$ that both of these upper bounds are realised.

Given $\psi \in \Gamma(\$)$, we may define $[\psi,\psi] \in \eX(M)$ by
transposing the Clifford action.  This is nonzero because this
bilinear product is now symmetric, since the inner product on $\$$ is
symplectic.  In fact, it is not difficult to show that $[\psi,\psi]$
is always causal; that is, it has non-positive minkowskian norm.

\begin{lemma}
  If $\psi \in \fg_1$ then $[\psi,\psi] \in \fg_0$.
\end{lemma}

\begin{proof}
  We first show that $[\psi,\psi]$ is Killing.  We first have that
  \begin{align*}
    g(\nabla_X[\psi,\psi],Y) &= X g([\psi,\psi],Y) -
    g([\psi,\psi],\nabla_X Y) && \text{since $\nabla g = 0$}\\
    &= X \left(\psi,Y\cdot \psi\right) - (\psi,\nabla_X Y \cdot \psi)
    && \text{by definition of $[\psi,\psi]$}\\
    &= \left(\nabla_X \psi,Y\cdot \psi\right) + \left( \psi,Y\cdot
      \nabla_X \psi\right)\\
    &= 2 \left(\nabla_X \psi,Y\cdot \psi\right) && \text{since
      $\left(\psi_1, Y \cdot \psi_2\right) = \left(\psi_2, Y \cdot
        \psi_1\right)$}\\
    &= -2 \left(Y \cdot \nabla_X \psi, \psi\right)~.
  \end{align*}
  Now since $D\psi = 0$,
  \begin{equation*}
    \nabla_X \psi = - \tfrac16 \iota_X F \cdot \psi - \tfrac1{12}
    X^\flat \wedge F \cdot \psi~,
  \end{equation*}
  whence
  \begin{align*}
    g(\nabla_X[\psi,\psi],Y) &= \tfrac13 \left( Y \cdot \iota_X F \cdot
      \psi, \psi\right) + \tfrac16  \left( Y\cdot (X^\flat \wedge F)
      \cdot \psi, \psi\right)\\
    &= \tfrac13 \left( (Y^\flat \wedge \iota_X F - \iota_Y \iota_X F) \cdot
      \psi, \psi\right)\\
    &\qquad + \tfrac16  \left( (Y^\flat \wedge X^\flat \wedge F +
      g(X,Y) F - X^\flat \wedge \iota_Y F) \cdot \psi, \psi\right)\\
    &= - \tfrac13 \left( \iota_Y \iota_X F \cdot \psi, \psi\right) +
    \tfrac16  \left( Y^\flat \wedge X^\flat \wedge F \cdot \psi,
      \psi\right)~,
  \end{align*}
  where we have used that for every 4-form $\Phi \in \Omega^4(M)$,
  \begin{equation*}
    \left(\Phi\cdot \psi, \psi\right) = 0~.
  \end{equation*}
  It follows that 
  \begin{equation*}
    g(\nabla_X[\psi,\psi],Y) + g(\nabla_Y[\psi,\psi],X)  = 0~,
  \end{equation*}
  whence $[\psi,\psi]$ is a Killing vector.  One can also prove that
  $\eL_{[\psi,\psi]} F = 0$.  Indeed, since $dF=0$, $\eL_{[\psi,\psi]}
  F = d\iota_{[\psi,\psi]} F$ and it is just a calculation to show
  that
  \begin{equation*}
    \iota_{[\psi,\psi]} F = - d B~,
  \end{equation*}
  where $B \in \Omega^2(M)$ is the 2-form in the square of $\psi$:
  \begin{equation*}
    B(X,Y) = \left(\psi, X^\flat \wedge Y^\flat \cdot \psi\right)~.
  \end{equation*}
\end{proof}

This result explains why $\psi$ is a called a Killing spinor, since
it is the ``square root''  of a Killing vector.  It follows by the
usual polarisation trick that if $\psi_1,\psi_2\in\fg_1$, then
$[\psi_1,\psi_2] \in \fg_0$.  We therefore have a symmetric bilinear
map $\fg_1 \times \fg_1 \to \fg_0$ denoted by $(\psi_1,\psi_2)
\mapsto [\psi_1,\psi_2]$.

We now define a bilinear map $\fg_0 \times \fg_1 \to \fg_1$ using the
spinorial Lie derivative $\eL_X$ defined in equation
\eqref{eq:Lieder}.  Let $X \in \fg_0$.  It follows from Proposition
\ref{pr:Lieder} and the fact that $X$ preserves $F$, that $\eL_X$
preserves $D$; that is,
\begin{equation}
  [\eL_X,D_Y] \psi = D_{[X,Y]} \psi\quad \forall~\psi \in
  \Gamma(\$)~,
\end{equation}
whence if $D\psi=0$, also $D\eL_X\psi = 0$.  Therefore $[X,\psi] =
\eL_X\psi$ defines the desired bilinear map.  Together with the Lie
bracket of vector fields, under which $\fg_0$ becomes a Lie algebra,
we have on $\fg = \fg_0 \oplus \fg_1$ the structure of a
superalgebra.

In checking the Jacobi identity, one again sees as in the first
lecture that 3/4 of the jacobator is identically zero because of
properties of the Lie derivative $\eL_X$.  The fourth component of the
jacobator vanishes if and only if for all $\psi \in \fg_1$,
\begin{equation}
  [[\psi,\psi],\psi] = 0 \qquad\text{or equivalently}\qquad
  \eL_{[\psi,\psi]} \psi = 0~.
\end{equation}
Representation theory is not useful here, since we are interested in a
general result for unspecified $\fg_0$ and $\fg_1$.  An explicit
calculation (made less painful with Mathematica or Maple) shows that
this is indeed the case.  Therefore we have a Lie superalgebra called
the \textbf{symmetry superalgebra} of the background.  The ideal
generated by $\fg_1$, $\fk = [\fg_1,\fg_1] \oplus \fg_1$ is called the
\textbf{Killing superalgebra} of the background.  Some (but not all,
see \cite{FHJMS-MESA}) backgrounds are such that their Killing
superalgebra admits an extension $\fm = \fm_0 \oplus \fg_1$ which is
``maximal''  in the sense that
\begin{equation*}
  \fm_0 = [\fg_1,\fg_1] \cong S^2\fg_1~,
\end{equation*}
where the isomorphism is one of vector spaces.  When it exists, it is
called the \textbf{maximal superalgebra} of the background.

\subsection{Examples}
\label{sec:examples-1}

Let us consider some examples.  The simplest is of course the
Minkowski maximally supersymmetric background $\RR^{1,10}$ with
$F=0$.  The symmetry superalgebra is the Poincaré superalgebra,
whereas the Killing superalgebra is the supertranslation ideal, as
explained in Section \ref{sec:elev-dimens-supergr}.  The maximal
superalgebra is obtained by taking $\fm_0 = S²\fg_1$ and declaring
$\fm_0$ to be central.  The extra elements in $\fm_0$ not in the
Killing superalgebra can be understood in terms of \emph{brane
  charges}, as explained, for example, in \cite{TownsendMTfS}.

Backgrounds with $F=0$ are said to be \emph{purely gravitational}.
The Killing spinors are parallel with respect to the Levi-Cività
connection $\nabla$.  This means that the holonomy of $\nabla$ is
contained in the stabiliser of a spinor in $\Spin(1,10)$.  There are
two types of spinor orbits in $S(1,10)$ and hence two stabilisers, up
to isomorphism.  As shown by Bryant \cite{Bryant-ricciflat} and the
author \cite{JMWaves}, the orbits are labelled by the value of a
quartic polynomial $q$, whose value $q(\psi)$ at $\psi \in S(1,10)$ is
the minkowskian norm of the vector $[\psi,\psi]$, which as mentioned
above is always non-negative.  If $q(\psi)=0$ we must distinguish
between the $\psi=0$ and a 25-dimensional orbit with stabiliser
isomorphic to $(\Spin(9)\ltimes \RR^8) \times \RR \subset
\Spin(1,10)$, whereas if $q(\psi) < 0$, the stabiliser is isomorphic
to $\SU(5)$.  This dichotomy gives rise to two types of supersymmetric
purely gravitational backgrounds: one generalising the M-wave
\cite{Mwave}, where $[\psi,\psi]$ is a lightlike parallel vector and
thus the geometry is described by a Brinkmann metric, and another
generalising the Kaluza--Klein monopole \cite{SMKK,GPMKK,HKMKK}.  This
latter class gives rise to reducible geometries of the form $\RR
\times N$, where $N$ is a riemannian ten-dimensional manifold with
holonomy contained in $\SU(5)$.  The Kaluza--Klein monopole is the
case $N = \RR^6 \times K$, with $K$ a 4-dimensional hyperkähler
manifold.  Because $\psi$ is parallel with respect to $\nabla$, so is
$[\psi,\psi]$ and the resulting Killing superalgebras are of the
supertranslation type, sketchily $[Q,Q] = P_+$, where $P_+$ is the
parallel null vector in the case of the waves, or else $[Q,Q] =$
translation in the flat factor, for the generalised Kaluza--Klein
monopoles.

For backgrounds with nonzero $F$, as in the M2-brane discussed in the
second lecture, the Killing superalgebra is still of the
supertranslation type, where now $[Q,Q] =$  translations along the
brane worldvolume.

For the maximally supersymmetric Freund--Rubin backgrounds the
symmetry superalgebra is simple and isomorphic to $\fosp(8|4)$ in the
case of $\text{AdS}_4 \times S^7$ and to $\fosp(6,2|2)$ for
$\text{AdS}_7 \times S^4$.  Simplicity implies that the Killing
superalgebra agrees with the symmetry superalgebra.  In
\cite{FHJMS-MESA} we showed via an explicit geometric construction
that the maximal superalgebra is isomorphic to $\fosp(1|32)$.

It was shown in \cite{Limits} that the Penrose limit contracts the
Killing superalgebra, whence for the maximally supersymmetric KG
background, obtained via the Penrose limit from the above
Freund--Rubin backgrounds,  the Killing superalgebra is a contraction
of either of the orthosymplectic superalgebras $\fosp(8|4)$ or
$\fosp(6,2|2)$.  This contraction was performed explicitly in
\cite{HatKamiSaka} obtaining the Killing superalgebra previously
computed in \cite{FOPflux}.

\subsection{The homogeneity conjecture}
\label{sec:homog-conj}

Because squaring Killing spinors one obtains Killing vectors, it
stands to reason that the more supersymmetric a background is, the
more isometries it has.  It is therefore tempting to conjecture that
there given sufficient supersymmetry --- that is, a sufficiently large
value of the fraction $\nu$ --- the background might be homogeneous,
where we say that a supergravity background $(M,g,F)$ is
\textbf{homogeneous} if it admits the transitive action of a Lie group
via $F$-preserving isometries.  So the question is whether there is
some critical fraction $\nu_c$ such that if a background has a
fraction $\nu > \nu_c$, then it is forced to be homogeneous.  All
maximally supersymmetric backgrounds discussed in the second lecture
are symmetric spaces, whence in particular homogeneous.  On the other
hand, the M2-brane, which has $\nu = \half$, has cohomogeneity one:
with orbits labelled by the radial coordinate $r$ in the solution.
This suggests that $\nu_c \geq \half$.  Furthermore, inspecting the
catalogue of known solution with $\nu > \half$, one sees that they are
always homogeneous.  This prompted Patrick Meessen to state the

\begin{HomConj}
  All supergravity backgrounds with $\nu>\half$ are homogeneous.
\end{HomConj}

In fact, we have to be a little careful because in practice we have
that $g,F$ are only locally defined in some open neighbourhood of
$\RR^{11}$, so that a more relevant notion is that of \emph{local
  homogeneity}, which is implied by \emph{local transitivity}, by
which we mean that around every point there is a local frame
consisting of $F$-preserving Killing vectors.

In \cite{FMPHom} we proved something weaker and at the same stronger
than the homogeneity conjecture.  We proved that if a background has
$\nu > \frac34$ then it is locally homogeneous, but we proved that
already $[\fg_1,\fg_1]$ acts locally transitively.  In other words,
local homogeneity is a direct consequence of supersymmetry.

\subsection{Status of the conjecture}
\label{sec:status-conjecture}

The conjecture does not just make reference to eleven-dimensional
supergravity, but in fact to any Poincaré supergravity theory.
Concentrating for definiteness on the ten-dimensional supergravity
theories, similar results exist for these theories as well.  In
\cite{EHJGMHom} we showed that any background of either type IIA or
IIB supergravity with $\nu > \frac34$ is locally homogeneous, whereas
any background of type I/heterotic supergravity with $\nu > \half$ is
locally homogeneous.  This latter result benefited from the
classification of parallelisable backgrounds \cite{FKYHeterotic},
which in turn was made possible by the fact that the Killing spinors
are defined by the lift to the spin bundle of a metric connection with
torsion.  We believe that the conjecture is true as stated, but
proving this for type II and eleven-dimensional supergravities will
require a better understanding of the connection $D$.

\section*{Acknowledgments}

These notes are a reasonably faithful transcription of three lectures
delivered at the Universidad Complutense de Madrid in November 2008 as
part of the \emph{Workshop on Higher Symmetries in Physics}.  I would
like to extend my gratitude to Marco Castrillón López for the
invitation.  The notes are an expanded version of two lectures given
at the Universität Bielefeld in May 2008 as part of the \emph{Meeting
  on Geometry and Supersymmetry} organised by Andriy Haydys, to whom I
am grateful for the invitation.

The results detailed in the first lecture had their origin in lectures
delivered in May 2007 at the Dipartamento di Matematica ``U. Dini'' of
the Università degli Studi di Firenze.  It is a pleasure to take this
opportunity to thank Dmitri Alekseevsky and Andrea Spiro for arranging
that visit, Luigi Mangiarotti for support, Cristina Giannotti for the
many delicious meals and Suor Tarcisia of \emph{le Suore ``Stabilite
  nella Carità'' della Villa Agape} for keeping me off the streets at
night.

Most of the work described here is based on collaborations with a
number of people, whom it is my pleasure to thank: Bobby Acharya,
Matthias Blau, Emily Hackett-Jones, Chris Hull, Patrick Meessen,
George Moutsopoulos, George Papadopoulos, Simon Philip, Hannu
Rajaniemi, Joan Simón and Bill Spence.

Finally, these notes were prepared while on sabbatical at the
Universitat de València, supported under research grant FIS2008-01980.
I am grateful to José de Azcárraga for making this visit possible.

\bibliographystyle{utphys}
\bibliography{AdS,AdS3,ESYM,Sugra,Geometry,Algebra}

\end{document}